\documentclass[twocolumn,trackchanges]{aastex631}
\accepted{}
\submitjournal{ApJ}
\shorttitle{dCCF analysis for MAXI J1820$+$070}
\shortauthors{Omama et al.}
\graphicspath{{./}{figs/}}

\usepackage{amsmath}
\begin{document}
\title{X-ray time lag evaluation of MAXI J1820+070 with a differential cross-correlation analysis} 

\author[0000-0002-4192-4424]{Tomoki Omama}
\affiliation{Institute of Space and Astronautical Science, Japan Aerospace Exploration Agency,\\
3-1-1 Yoshinodai, Chuo-ku, Sagamihara, Kanagawa 252-5210, Japan}
\affiliation{Department of Space and Astronautical Science, School of Physical Science, The Graduate University for Advanced Studies, SOKENDI \\
3-1-1 Yoshinodai, Chuo-ku, Sagamihara, Kanagawa 252-5210, Japan}

\author[0000-0002-9184-5556]{Masahiro Tsujimoto}
\affiliation{Institute of Space and Astronautical Science, Japan Aerospace Exploration Agency,\\
3-1-1 Yoshinodai, Chuo-ku, Sagamihara, Kanagawa 252-5210, Japan}
\affiliation{Department of Space and Astronautical Science, School of Physical Science, The Graduate University for Advanced Studies, SOKENDI \\
3-1-1 Yoshinodai, Chuo-ku, Sagamihara, Kanagawa 252-5210, Japan}

\author[0000-0002-5352-7178]{Ken Ebisawa}
\affiliation{Institute of Space and Astronautical Science, Japan Aerospace Exploration Agency,\\
3-1-1 Yoshinodai, Chuo-ku, Sagamihara, Kanagawa 252-5210, Japan}
\affiliation{Department of Astronomy, Graduate School of Science, The University of Tokyo \\
7-3-1 Hongo, Bunkyo-ku, Tokyo 113-8654 Japan}

\author[0000-0003-2161-0361]{Misaki Mizumoto}
\affil{Hakubi Center, Kyoto University, Yoshida-honmachi, Sakyo-ku, Kyoto, 606-8501, Japan}
\affil{Department of Astronomy, Graduate School of Science, Kyoto University, Kitashirakawa-oiwakecho, Sakyo-ku, Kyoto, 606-8502, Japan}

\begin{abstract}
 MAXI J1820$+$070 is a transient black hole binary (BHB) discovered on 2018 March 11.
 The unprecedented rich statistics brought by the NICER X-ray telescope
 allows detailed timing analysis up to $\sim$1~kHz uncompromised by the photon shot noise.
 To estimate the time lags, the Fourier analysis was applied,
 which led to two different conclusions for the system configuration; one supporting
 the lamp-post
 configuration with a stable accretion disk extending close to the innermost stable
 circular orbit and the other supporting the truncated accretion disk contracting with
 time. Using the same data set, we present the results based on the cross-correlation
 function (CCF).
 The CCF is calculated between two different X-ray bands and one side
 is subtracted from the other side,
 which we call the differential CCF (dCCF).
 The soft and hard lags respectively of $\sim$0.03 and 3~s
 are clearly identified without being diluted by the spectral mixture, demonstrating
 the effectiveness of the dCCF analysis. The evolution of these lags is tracked, along
 with spectral changes for the first 120~days since the discovery.
 Both the dCCF and spectral fitting results are interpreted that the soft lag is a
 reverberation lag between the Comptonized emission and the soft excess emission and that
 the hard lag is between the disk black body emission and the Comptonized emission. The
 evolution of these lags is in line with the picture of the truncated disk contracting
 with time.
\end{abstract}

\keywords{Stellar mass black holes (1611) --- X-ray astronomy(1810) --- X-ray transient
sources(1852) --- Reverberation mapping(2019)}

\section{Introduction}
\label{sec:intro}

MAXI J1820+070 (hereafter J1820) is a transient black-hole binary (BHB) that exhibited
outbursts four times in a row from March 2018 to October 2019. It was first reported in
the optical band by ASSASSN \citep{tucker_asassn-18ey_2018} and observed in the X-ray
band by the Monitor of All-sky X-ray Image (MAXI) \citep{matsuoka_maxi_2009} a few days 
later. Due to its extreme brightness (4~Crab at the maximum; \citealt{Shidatsu2019}) and
low galactic absorption ($N_H \sim 10^{-21}~\text{cm}^{-2}$; \citealt{uttley_nicer_2018})
at a short distance ($3\pm 1$~kpc; \citealt{gandhi_gaia_2019}), J1820 was intensively 
observed with a suite of observatories and was found to exhibit many behaviors similar
to other BHBs. The Neutron star Interior Composition Explorer (NICER;
\citealt{gendreau_neutron_2016}) observed J1820 most frequently, providing a unique data
set that enables detailed X-ray timing-spectroscopy analyses up to a $\sim$1~kHz range
against the photon shot noise.

The combination of the favorable features of J1820 and the rich data set of
NICER are ideal to investigate some of the unsolved problems of BHBs. One of
them is the geometry of the hot corona in the low/hard state, which is considered to be
the production site of the Comptonized emission that dominates the X-ray band. Two
competing pictures are proposed; one is the hot accretion flow and truncated-disk model
\citep{plant_revealing_2014,ingram_physical_2011}, in which the optically-thick disk is 
truncated at a
radius far away from the innermost stable circular orbit (ISCO) from which matter falls
along the advection-dominated accretion flow (ADAF). Thermal photons are thought to come
from the optically-thick disk, which are Comptonized in the hot corona in the ADAF.
The other is the lamp-post model with a different configuration
\citep{fabian_determination_2014,garcia_x-ray_2015}; the accretion disk is close to the
ISCO radius and a small hot corona exists along the rotation axis of the black hole.

The above two models have been pursued for J1820. \citet{Kara2019} found that the
reverberation time lag from the Comptonized emission to its disk reflected emission is
shorter than that observed in other BHBs. Together with a stable Fe K$\alpha$ line
profile produced by reprocessing the Comptonized emission off of the disk, they
supported the lamppost configuration \citep{Kara2019,Buisson2019}. This
interpretation was challenged by many others using the same data set with various
techniques
\citep{DeMarco2021,Axelsson2021,Zdziarski2021,kawamura_full_2022,kawamura_maxi_2022},
all of which point to the truncated-disk model with the inner disk radius
much larger than the ISCO.

Amongst many analysis techniques, the reverberation time lag is a key for providing a
direct constraint on the geometry. It is conventionally analyzed by using the cross 
spectrum in the frequency domain. Curiously, the two camps used this technique
with the same data set, but reached different conclusions. On one hand,
\citet{Kara2019} reported a shorter reverberation lag of $\mathcal{O}$(1~ms) using
the phase shift in the cross spectrum and argued for the lamppost geometry. On the other
hand, \citet{DeMarco2021} reported a longer lag of $\mathcal{O}$(10~ms) using the
zero-crossing points in the cross spectrum and argued for the truncated-disk geometry.

The time lags derived from the cross-spectrum need some caution, regardless of whether
the estimate is based on the phase shift or the zero crossing points, as we will
describe in \S~\ref{sec:method}. Here, we propose an approach based on
the cross-correlation function (CCF) as an alternative method to determine the time
lag. This is a time-domain technique complementary to the cross-spectrum and free from
the cautions associated with the frequency-domain technique. The CCF has not been an
often-used approach within X-ray bands since its early applications
\citep[e.g.][]{priedhorsky_extended-bandwidth_1979,maccarone_time_2000} mainly because
the small contrast between the two spectral components with a time lag yields a small
cross correlation that is overwhelmed by an autocorrelation of the component of
interest. However, with the availability of the NICER data with rich photon statistics
and a subtraction of the autocorrelation contributions, we revisit the utility of this
technique in actual applications to J1820. In addition to the CCF-based time lag, we
also perform spectral analysis to confirm that the CCF-based time lag result can be
interpreted in a picture consistently with the spectral analysis result.

\medskip

The plan of the paper is as follows. In \S~\ref{sec:method}, we give a formalism of the
CCF and compare it with the cross-spectrum. In
\S~\ref{sec:observations_and_data-reduction}, we describe the NICER data, to
which we apply the CCF. In \S~\ref{sec:analysis}, we start with an overview of the
data set (\S~\ref{subsec:2018-3-21}). We first apply the CCF and spectral analysis at
one epoch (\S~\ref{subsec:oneobs}) and expand the function to the entire
range of interest
(\S~\ref{sec:ccf_time-development}) to track the development of the CCF parameters. In
\S~\ref{sec:discussion}, we discuss interpretation of the result. We show that both the
CCF-based time lag and the spectral results can be interpreted with the truncated-disk
model with contracting inner radius as the low/hard state proceeds up till the
transition to the soft state. The summary and conclusion are given in
\S~\ref{sec:conclusion}.

\section{Method} \label{sec:method}
We briefly compare two approaches for the time lag analysis; one is the
conventionally-used cross spectra in the frequency domain and the other is the CCF in
the time domain used in this paper. Readers are referred to
\citet{poutanenImpactReverberationFlared2002} for derivation of the equations for a
typical disk geometry. We give a brief derivation below for the sake of completeness.

We take two X-ray light curves in the different energy bands 1 and 2. We
assume that the band 1 is dominated by the direct emission $x(t)$, whereas the band 2 is
a linear composition of the direct emission and its reprocessed emission lagged from the
direct emission by a fixed time $t_{\mathrm{lag}}$. The light curve of each band is
given by
\begin{equation}
 \begin{array}{ccl}
 y_1(t) &=& x(t) \\
 y_2(t) &=& \alpha x(t) + \beta x(t-t_{\mathrm{lag}}),
  \label{eq:lcurves}
 \end{array}
\end{equation}
where $\alpha$ and $\beta$ are linear coefficients. 

The cross spectrum, on one hand, is calculated in the frequency domain as
\begin{equation}
 \begin{split}
 S_{1,2}(f) &= Y_1^*(f) Y_2(f) \\
            &= \overline{x(t)^2} \left[
            \alpha + \beta e^{-i2\pi f t_{\mathrm{lag}}}
            \right],\\
 \end{split}
\end{equation}
in which $Y_{i}(f)$ denotes the Fourier transform of $y_{i}(t)$ for $i=1, 2$,
and ${\overline{x(t)^2}}$ is the time average for $x(t)^2$.
The argument of $S_{1,2}$ is
translated as the lag amplitude by
\begin{equation}
 \Delta t(f) = \frac{1}{2 \pi f} \arctan\left({\frac{\beta \sin{(2 \pi f t_{\mathrm{lag}})}}{\alpha + \beta \cos{(2 \pi f t_{\mathrm{lag}})}}}\right).
 \label{eq:timelag_cs}
\end{equation}
Only when $\alpha=0$, i.e., the band 2 is composed only of the reprocessed emission,
$\Delta t$ matches $t_{\mathrm{lag}}$. At the zero frequency limit,
\begin{equation}
 \Delta t(f) \rightarrow \frac{\beta}{\alpha + \beta} t_{\mathrm{lag}}~~(f \rightarrow 0).
\end{equation}
In many cases, the reprocessed emission is subtle compared to the direct emission within
the X-ray band ($\alpha \gg \beta$), thus the $t_{\mathrm{lag}}$ is diluted in $\Delta
t(f)$ by a large factor ($\frac{\beta}{\alpha + \beta}$)
\citep[][]{Miller2010}. Therefore, the face value of the lag amplitude in the
cross spectrum should not be used to derive the reverberation distance. The first
zero-crossing point of $\Delta t(f) =0$ at $f = \frac{1}{2 t_{\mathrm{lag}}}$ is a more
robust estimate of the lag amplitude \citep[][]{Mizumoto2018a}. In practice, though, it
is not easy to track the oscillatory behavior of $\Delta t(f)$ to identify the first
zero-crossing point in a small dynamic range of $f$ limited by the observation duration
and the time binning against the photon shot noise.

The CCF, on the other hand, is defined as the inverse Fourier transform of $S_{1,2}(f)$
in the time domain.  The real part is expressed as 
\begin{equation}
\begin{split}
  C(\tau) &= \Re \left\{
        \int_{-\infty}^{\infty} S_{1, 2} (f) e^{i2\pi f \tau} dt\right\} \\
      &= \alpha A(\tau) + \beta A(\tau-t_{\mathrm{lag}}), \\
\end{split}
 \label{eq:ccf_from_cs}
\end{equation}
where $A(\tau)$ is the auto-correlation function (ACF). Two peaks at $\tau=0$ and
$t_{\mathrm{lag}}$ appear separately with an amplitude contrast of $\alpha/\beta$.

The cross spectra have been more dominantly used to date in assessing the reverberation
time lags in the X-ray band \citep{Uttley2014}. This is
because the contrast $\beta/\alpha$ is usually too small within the X-ray bands.
Only when high contrast is expected between two different wavelengths, the
CCF becomes a powerful technique. This was indeed the case in J1820 between X-ray and
UV or optical bands \citep{Kajava2019,Paice2021}. Now, when combined with rich
photon statistics that NICER brings for the first time, the CCF, or a variant
thereof, can be used between two different X-ray bands even for a small
contrast. We will demonstrate this by using the NICER data of J1820.

\section{Observations and data reduction} \label{sec:observations_and_data-reduction}
We use the data obtained with NICER, with which J1820 was observed many
times. The entire view of the NICER data set can be found in
\citet{Stiele2020}. We use the period from MJD~58190 to 58304, which covers the
first brightening followed by the gradual decline and the onset of the second
brightening. This period has been intensively studied by several authors, leading to
the dispute of its interpretation described in \S~\ref{sec:intro}.

NICER is an X-ray timing-spectroscopy mission operated in the international
space station since 2017. 
It consists of 56 independent X-ray Timing Instruments (XTI), each of which has the
X-ray concentrator optics and the silicon drift detector
\citep{gendreau_neutron_2016}. In addition to its specific mission goals
\citep{gendreau_neutron_2016}, NICER is useful in a wide range of astrophysics.
Its large effective area and large dynamic range allow timing analysis beyond 1 kHz,
which has been inaccessible with other X-ray instruments due to photon shot noise or
slow response. Moreover, its moderate energy resolution (140~eV at 6~keV) and bandpass
(0.5--12~keV) make broad-band spectral analyses possible. For the purpose of this study,
NICER is uniquely suited.

We retrieved the pipeline products from the archive for the observation sequences
1200120101--1200120196. Detailed information on the data set can be found in
\citet{DeMarco2021}. We used HEASoft version 6.29 for data analysis. We generated the
redistribution matrix function (RMF) and the auxiliary response function (ARF) for each
data set. We used the background file provided by the instrument team.

\section{Analysis} \label{sec:analysis}
We start with a brief overview of the data set
(\S~\ref{subsec:2018-3-21}). We first use the data at one epoch to
describe the analysis in detail (\S~\ref{subsec:oneobs}), which will be expanded in the
entire duration (\S~\ref{sec:ccf_time-development}).

\subsection{Overview of data} \label{subsec:2018-3-21}
After the first X-ray detection of J1820 \citep{kawamuro_maxigsc_2018} on March 11, 2018,
the X-ray flux increased rapidly for two weeks to reach the peak flux of $\sim$4~Crab on
March 26. The spectral hardness remained hard and the flux gradually decreased for the
next $\sim$90 days (the low hard state in \citealt{Shidatsu2019}) when the source made
a drastic transition to the high soft state. The source exhibited further changes later,
which is out of the scope of this paper. We focus on the entirety of the low hard state
and the onset of the transition to the high soft state, in which a dense coverage by
NICER was made for $\sim$120 days. The detailed description of the data set is
given extensively in the literature
\citep{Kara2019,Stiele2020,Axelsson2021,DeMarco2021}. Here, we
only give two characterizations ---the light curve and the color-color diagram--- for the
sake of completeness of this paper.

\begin{figure}
 \plotone{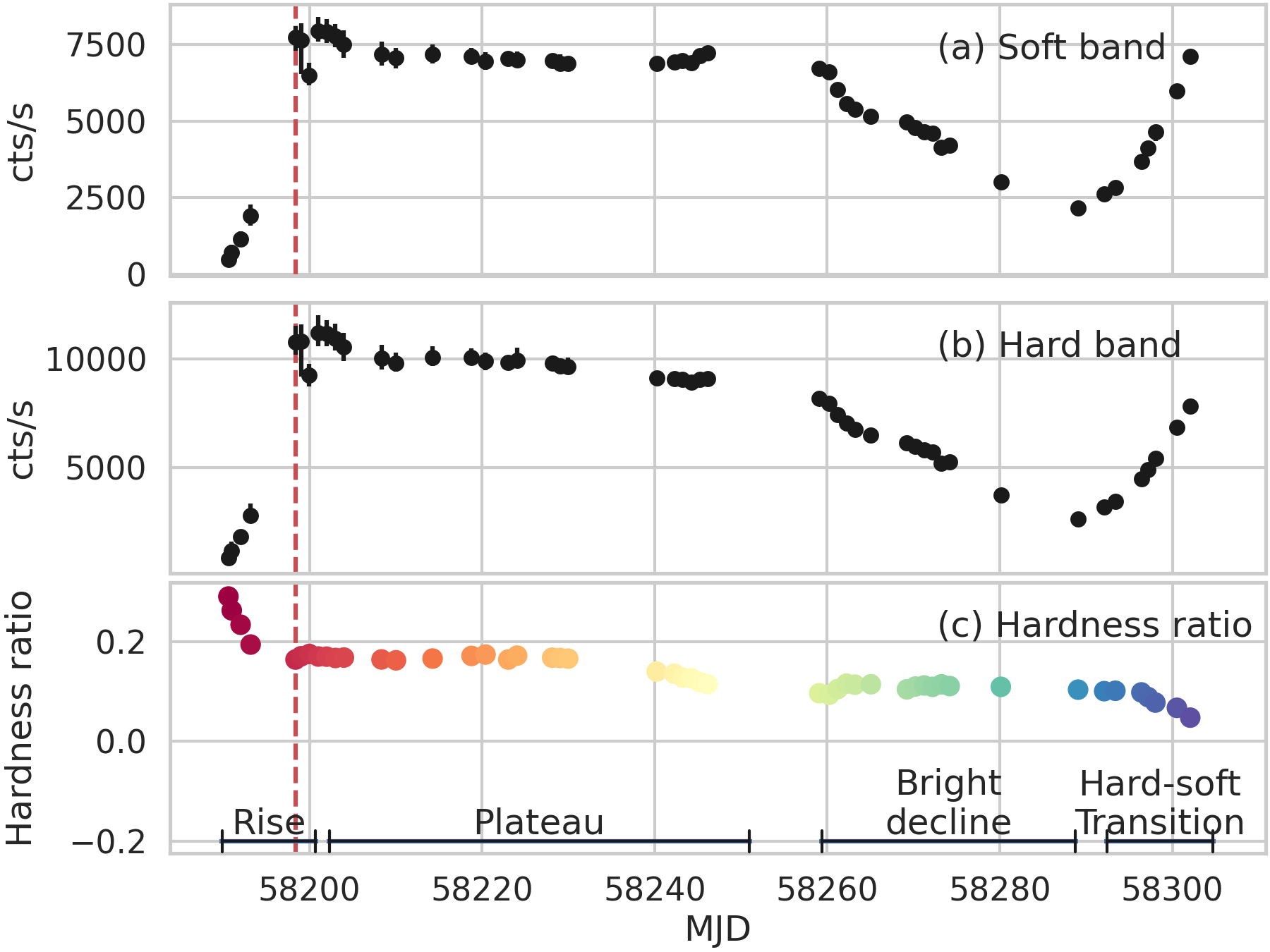}
 \caption{X-ray light curves in the soft (0.5--1 keV) and hard (1--10 keV) bands (top
 and middle panels, respectively) and the hardness ratio (bottom panel). Each point is
 calculated for a mean value among 64~s bins of the light curve in a certain
 duration. Colors in the hardness ratio represent MJD in the same manner as in
 Figure~\ref{fig:CCDIAG_first-hard-state}. The phase definitions are given in the bottom
 panel. The red vertical dashed line shows MJD 58197 (2018 March 21) used in the
 detailed analysis (\S~\ref{subsec:oneobs}).}
 \label{fig:TS-COUNT_first-hard-state}
\end{figure}

Figure~\ref{fig:TS-COUNT_first-hard-state} shows the X-ray light curve in the soft
(0.5--1.0~keV) and hard (1.0--10.0~keV) bands ($S(t)$ and $H(t)$, respectively) and the
spectral hardness defined as $\left\{H(t)-S(t)\right\}/\left\{H(t)+S(t)\right\}$. After
the rapid brightening at the beginning, the flux change was slow with a noticeable break
between MJD 58240 and 58260. In the last part after MJD 58280, the source started a
transition to a new state, which is dubbed as the intermediate state
\citep{Shidatsu2019}. Hereafter, we follow the phase definition by \citet{DeMarco2021}:
i.e., rise (until MJD 58201), plateau (until MJD 58251), bright decline (until MJD 58289),
and hard-soft transition (until MJD 58305). 

\begin{figure}
\plotone{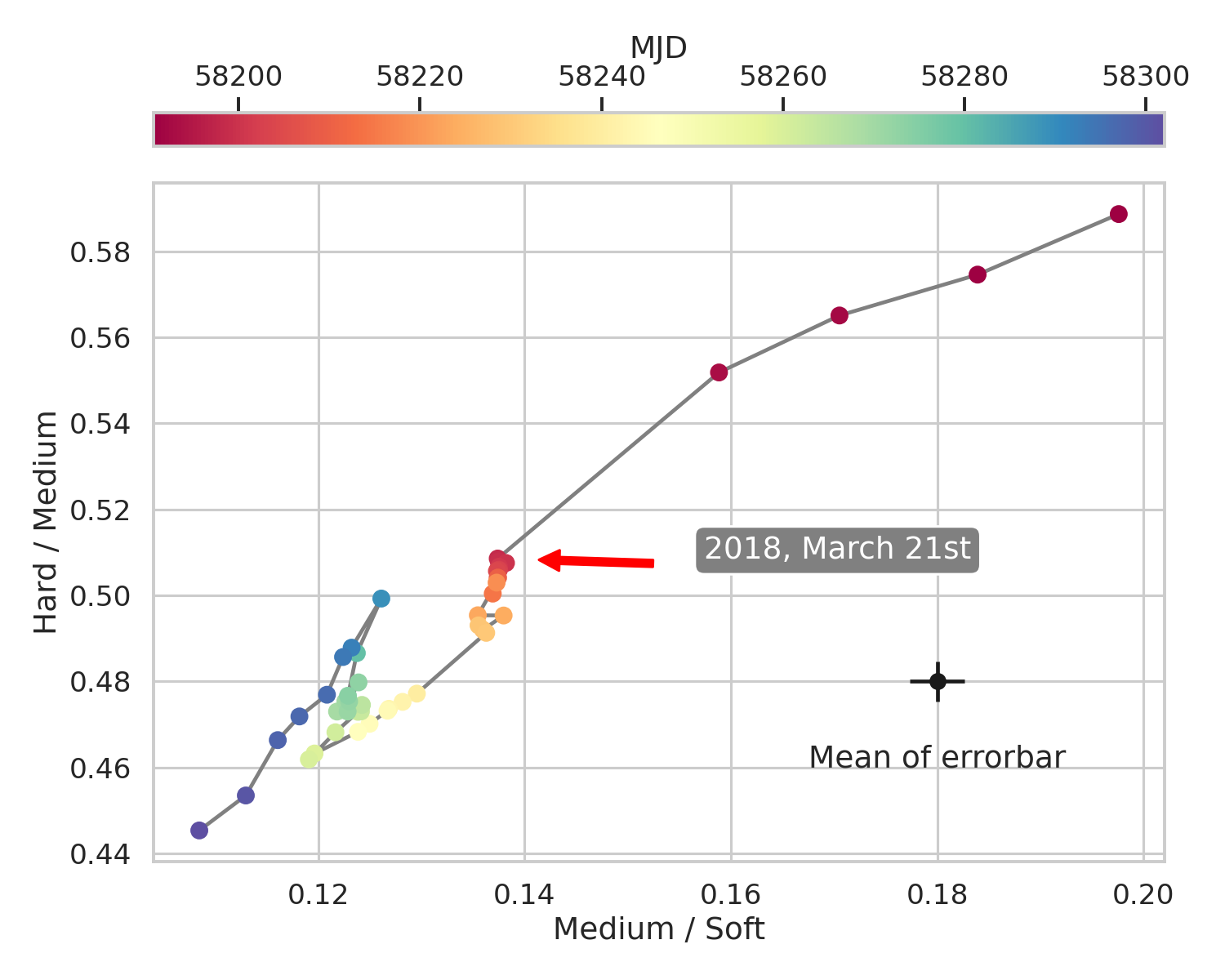}
 \caption{X-ray color-color diagram. Here, we defined the soft, medium, and hard bands as 0.5--2,
 2.0--3.5, and 3.5-10.0~keV, respectively. Two colors are calculated as the ratio of the 
 two adjacent bands. The black curve gives the track connecting data points. MJD 58197
 (2018 March 21) is shown by the red arrow, which is used for the detailed analysis
 (\S~\ref{subsec:oneobs}).}
 \label{fig:CCDIAG_first-hard-state}
\end{figure}

Figure~\ref{fig:CCDIAG_first-hard-state} shows evolution of the spectral change. Three
bands are defined as soft (0.5--2~keV), medium (2.0--3.5~keV), and hard (3.5--10~keV)
and the ratios of the adjacent two bands are plotted. The color codes are given to aid
chronological interpretation. The initial rapid change (red), the gradual and monotonic
decay with a break between MJD 58240 and 58260 (yellow), and the start of the state
transition (green) are clearly traced.

\subsection{Observation of 2018 March 21} \label{subsec:oneobs}
We use the data (OBSID 1200120106) at the peak flux on 2018 March 21 (MJD 58197) to
establish the data analysis method. We first fit the X-ray spectrum with a
phenomenological model to decompose it into major spectral components and select energy
bands where they are most dominant (\S~\ref{subsec:spectrum_fit_1200120106}). We then
apply CCF analysis between these bands (\S~\ref{ssubsec:ccf_oneobs}).

\subsubsection{Spectral decomposition}\label{subsec:spectrum_fit_1200120106}
A phenomenological model fitting was performed in 0.7--10~keV to define appropriate
energy bands for the CCF analysis. In the hard state, the Comptonized component in the
power-law form is dominant in the entire energy range \citep{Axelsson2021}. Deviations
are seen in the soft band with another broad component attributable to the disk
emission, and in the Fe K band with both narrow and broad emission lines
\citep{Axelsson2021}. In addition, an excess emission in the softest band is required
for complete spectral modeling \citep{DeMarco2021}.

We include all these components to fit the spectrum. The multi-color disk black body
component was represented using the \texttt{diskbb} model in the \texttt{XSpec} spectral
fitting package. A part of the disk emission is inverse Compton scattered into
power-law emission, which was modeled with the \texttt{simpl} model. The soft excess
component was represented by a broad Gaussian component. The narrow and broad lines at
the Fe K band are the Gaussian components with their center fixed at 6.4~keV. The
summation is attenuated by an interstellar extinction by \texttt{TBabs}
\citep{wilms_absorption_2000} with the hydrogen-equivalent column fixed at $1.5 \times
10^{21}$~cm$^{-2}$ \citep{uttley_nicer_2018}.

\begin{figure}
 \plotone{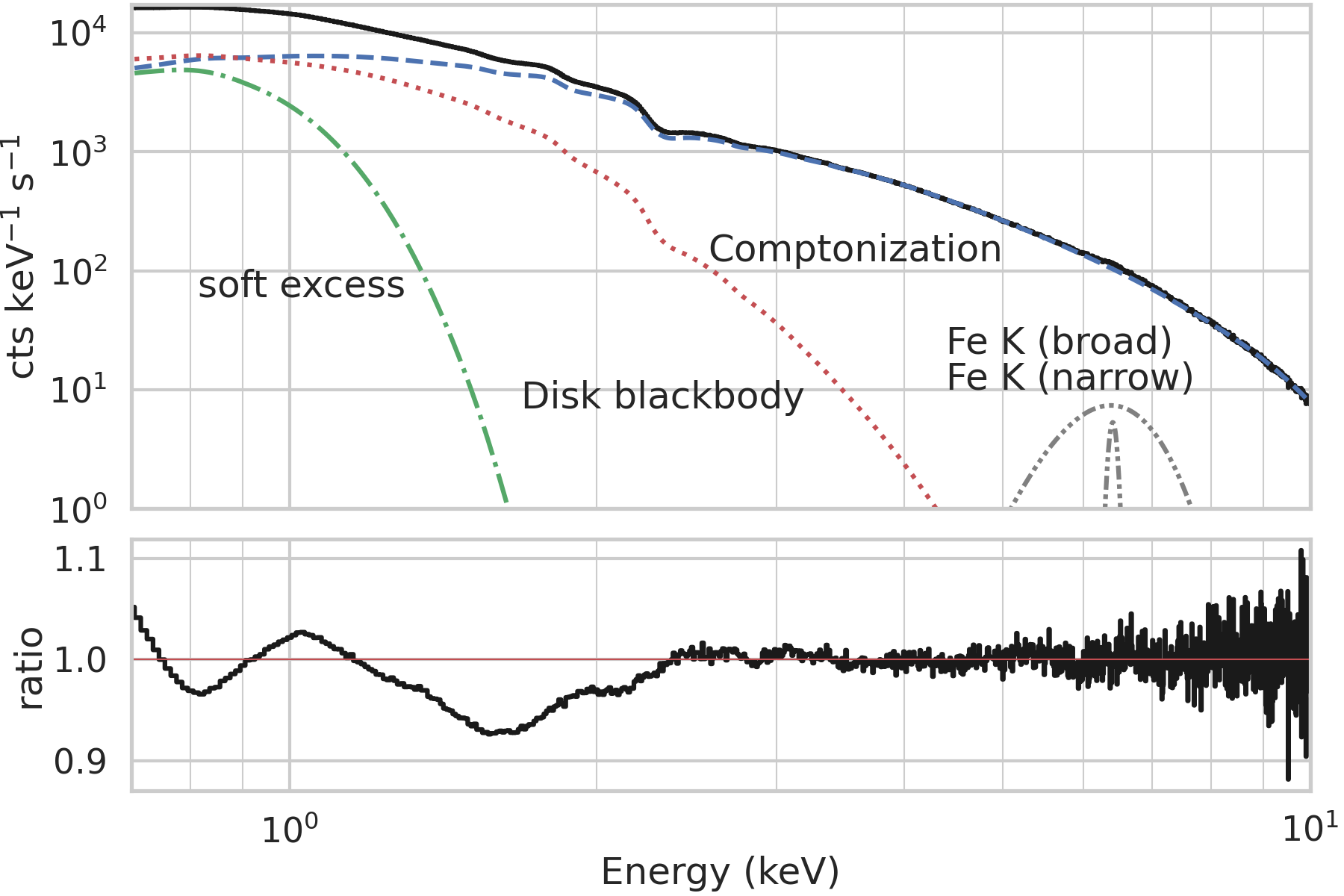}
 \caption{Observed spectrum on MJD58197 and the best-fit model (top), and residuals of the fit
 (bottom). Models of each spectral component are shown with different symbols.}
\label{fig:XSPEC-FIT_1200120106}
\end{figure}

The three broad components (disk black body, Comptonized, and soft excess) are coupled
with each other at the softest band in the spectral fitting. In order to decouple them
as much as possible, we first fitted the spectrum in the 2.2--10.0~keV band where the
soft excess emission is negligible, and constrained the disk black body and its
Comptonized components along with the Fe K emission lines. We then extrapolated the
best-fit model to the entire 0.7--10.0~keV band and added the soft excess component.
The spectrum, the best-fit model, and the residuals to the fit are shown in
Figure~\ref{fig:XSPEC-FIT_1200120106}. The detector response calibration and the fitting
models are presumably imperfect at this moment, leaving some structures below
$\sim$3~keV in the residual, which we ignore in this paper, because the purpose here is
to define the appropriate energy bands for the following CCF analysis.

The Comptonized component dominates the entire energy range. The disk emission is
sub-dominant below $\sim$5 keV and the soft excess emission below $\sim$2~keV. We thus
define five fine energy bands: 0.5--1, 1--2, 2--3, 3--5, and 5--10~keV having different
mixtures of these spectral components. Alternatively, we also use the two coarse bands,
the soft band (0.5--1~keV) and the hard band (1--10~keV).

\subsubsection{CCF} \label{ssubsec:ccf_oneobs}
\begin{figure*}
 \plotone{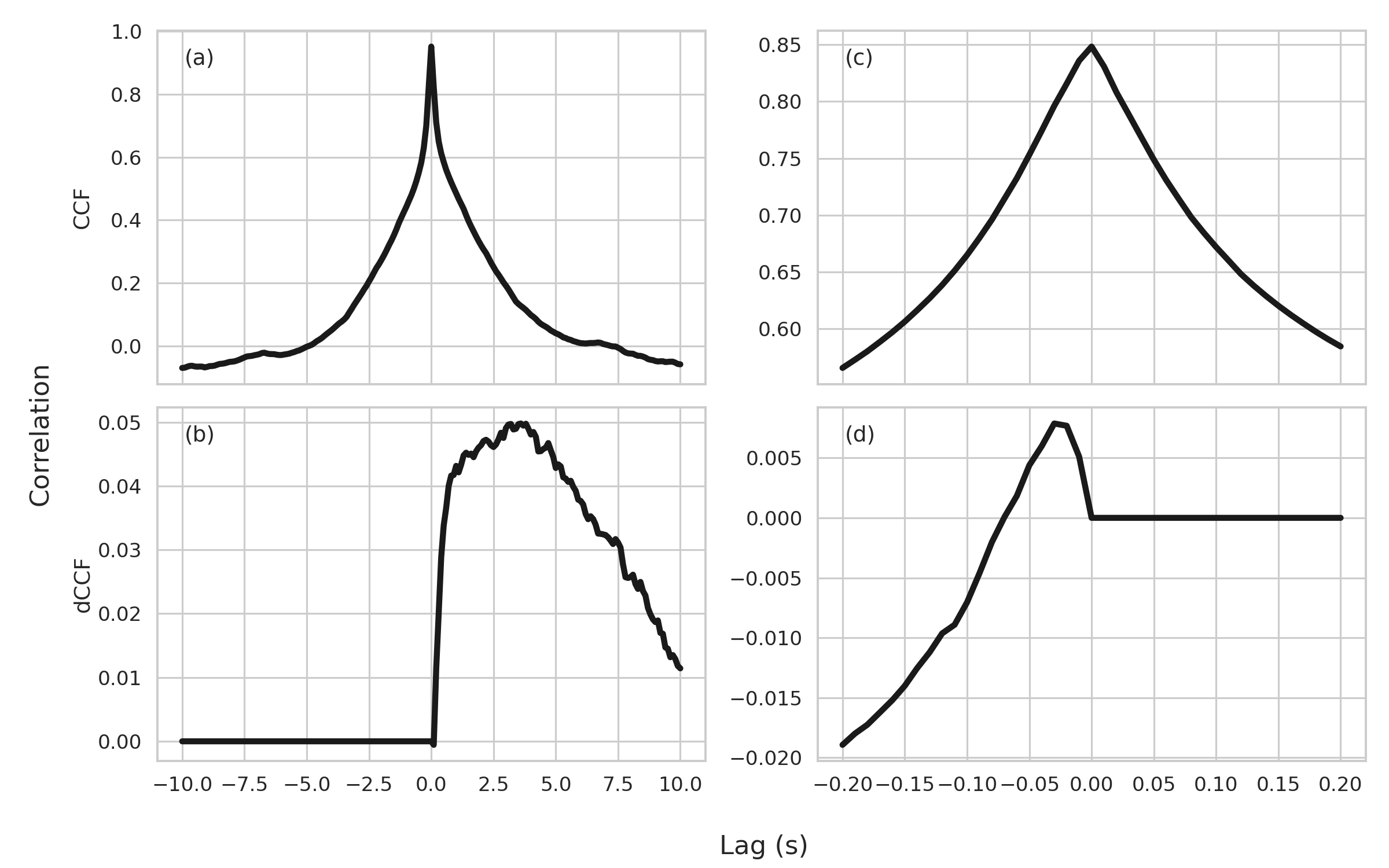}
 \caption{(top) CCF and (bottom) dCCF between the two coarse energy bands ---soft
 (0.5--1.0~keV) and hard (1.0--10.0~keV) bands--- in the longer ($\tau < 10$~s; left)
 and the shorter ($\tau < 0.2$~s; right) time scales.}
 \label{fig:CCFDCCF_05t1-1t10_1200120106}
\end{figure*}

We calculated the CCF between the two coarse bands (soft and hard), and show the result
in two different time scales: (1)
$-10 \le \tau \le +10$~s (panel a) and (2) $-0.2 \le \tau \le + 0.2$~s (panel c) in
Figure~\ref{fig:CCFDCCF_05t1-1t10_1200120106}. Because both bands are dominated by the
same spectral component (Comptonized component), the CCF is dominated by its auto
correlation peaking at 0~s. However, there are noticeable asymmetry in both time scales,
which are clarified by subtracting the left half ($\tau < 0$~s) from the right half
($\tau>0$~s) or vice versa as shown in panels (b) and (d). The subtraction sign was
swapped arbitrarily to make the lag to be positive; i.e., we interpret the negative soft
lag as positive hard lag. We call them differential CCF (dCCF).

\begin{figure*}
 \plotone{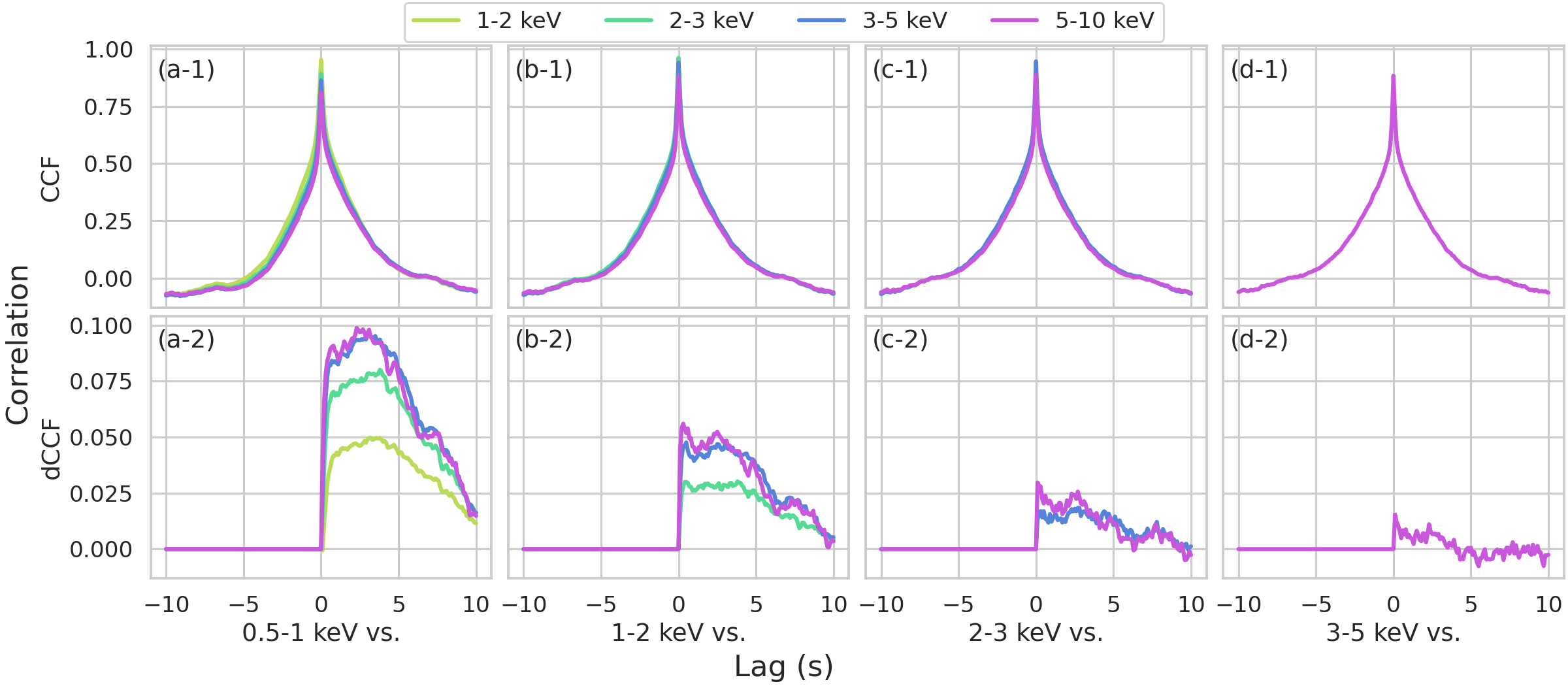}
 \caption{CCF (top) and dCCF (bottom) between two of the five fine energy bands in the
 longer time scale of $|\tau| \le 10$~s with a time bin of 0.1~s. The reference band is
 given at the bottom of each column, while the target band is shown with different
 colors.}
\label{fig:CCFDCCF-MULTI_05t1-spec_1200120106_hardlag}
\end{figure*}

\begin{figure*}
 \plotone{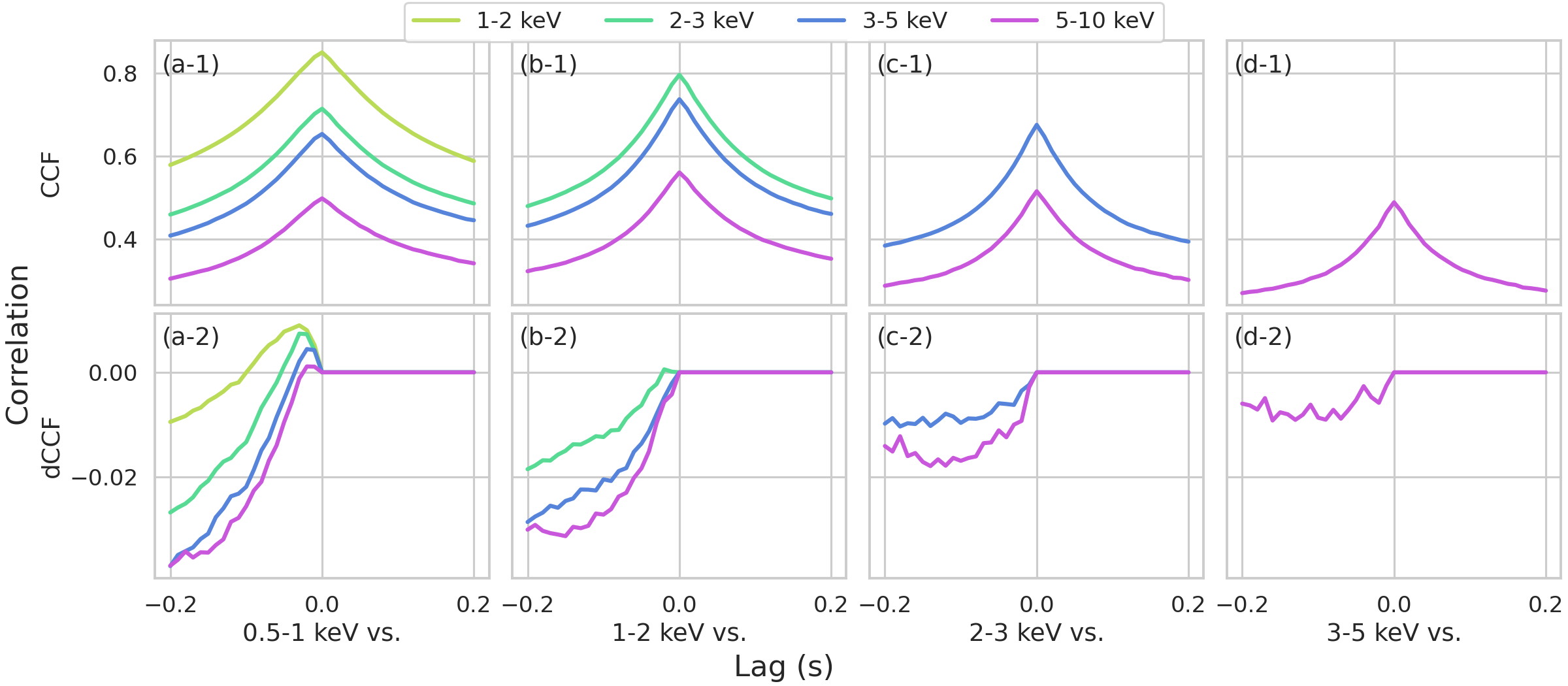}
\caption{Same as Figure~\ref{fig:CCFDCCF-MULTI_05t1-spec_1200120106_hardlag} for the
 shorter time scale of $|\tau| \le 0.2$~s with a time bin of 0.01~s.}
 \label{fig:CCFDCCF-MULTI_05t1-spec_1200120106_softlag}
\end{figure*}

Two distinct structures are found in the dCCF. One is a positive hard lag in the longer
time scale (panel b) and the other is a positive soft lag in the shorter time scale
(panel d). These lag phenomena correspond to the hard and soft lags found using the
cross spectra \citep{Kara2019,DeMarco2021}, where their amplitudes were 3~s and
30~ms, respectively. We emphasize that the same amount of the lags free from the
spectral dilution are easily recognized in the dCCF. Furthermore, the two lag amplitudes
are not a single value (not a delta function) but have some structures in the dCCF,
which may carry information on the lag response.

Next, we use the finer energy bands to investigate energy dependence of these lags.
Two energy bands out of the five were selected as the reference and the target band
for the CCF. Figures~\ref{fig:CCFDCCF-MULTI_05t1-spec_1200120106_hardlag} and
\ref{fig:CCFDCCF-MULTI_05t1-spec_1200120106_softlag} respectively show the hard lag
in the longer time scale and the soft lag in the shorter time scale, where 
the CCF (top panels) and the dCCF (bottom panels) between different bands are
calculated. Note that the dCCF for the soft lag is affected by the one
for the stronger hard lag with a longer time scale. Still, we see positive
soft lag clearly in the dCCF at panel (a-2) in
Figure~\ref{fig:CCFDCCF-MULTI_05t1-spec_1200120106_softlag}.

\subsection{Development of the spectral and temporal behaviors} \label{sec:ccf_time-development}

\begin{figure*}
 \epsscale{0.90}
 \plotone{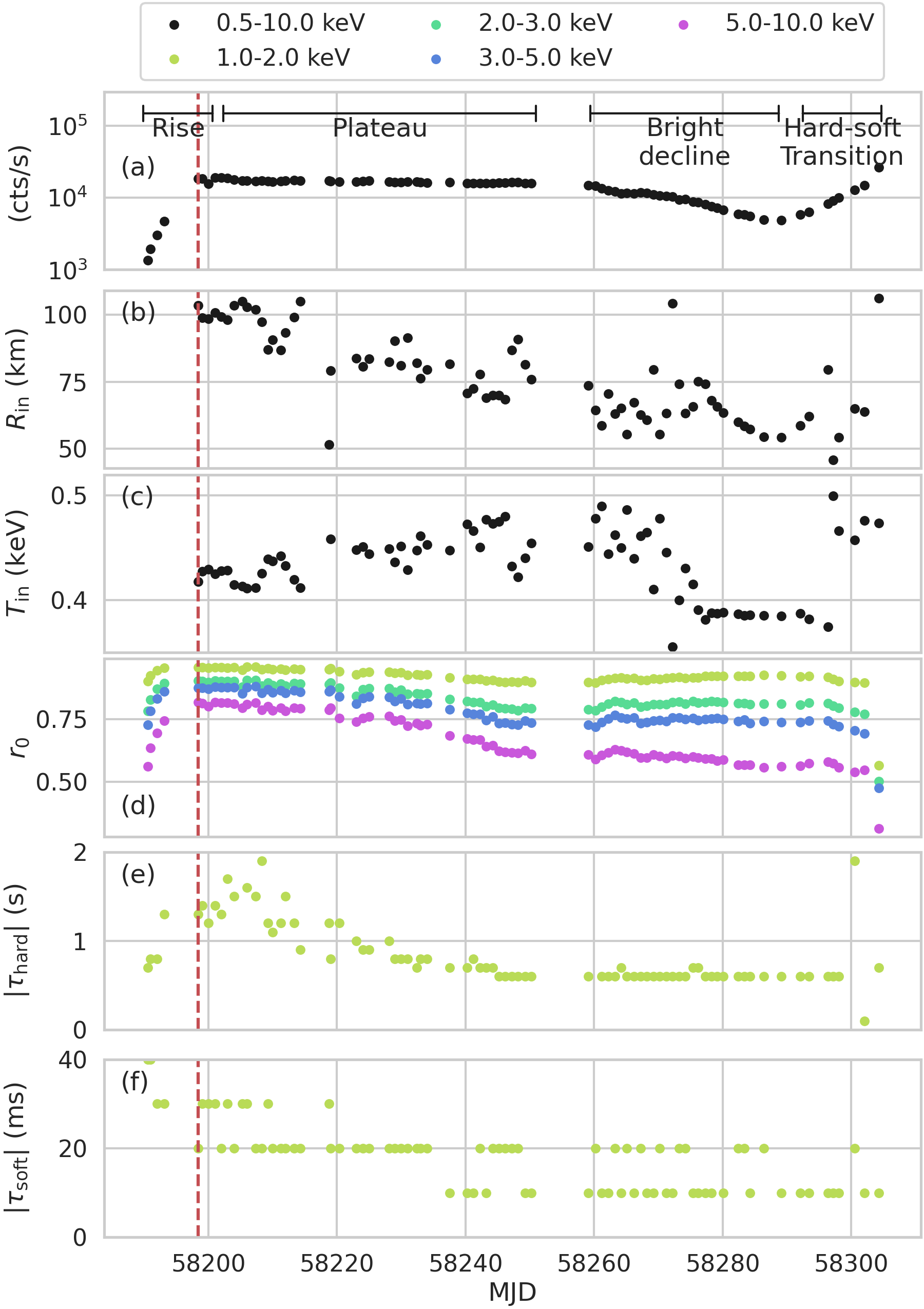}
 \caption{Development of (a) the count rate in the 0.5-10.0~keV band, (b--c) the
 spectral parameters ($R_{\mathrm{in}}$ and $T_{\mathrm{in}}$ in the
 \texttt{diskbb*simpl} model fitting, and (d--f) the CCF and dCCF parameters; the
 zero-lag correlation in CCF (d), the mean hard and soft lags in (e and f,
 respectively). The CCF and dCCF for only 0.5-1.0~keV and 1.0-2.0~keV are shown.
 We plot the error bars in the (b) and (c), which are hardly seen
 due to the small values. The red vertical dashed line shows MJD~58197 (2018 March 21)
 used in the detailed analysis \S~\ref{subsec:2018-3-21}. Pairs of the same energy band
 in all figures use the same color shown in the legend.}
 \label{fig:ts_parameters}
\end{figure*}

We apply the spectral and temporal analyses developed in one epoch
(\S~\ref{subsec:oneobs}) to the entire data set. The same analysis worked for the most
part of the changing behavior J1820 in the duration of
interest. Figure~\ref{fig:ts_parameters} shows the development of the count rate in
panel (a) along with the results of the spectral analysis in the panels (b--c) and those
of the temporal analysis in the panels (d--f). The phase definition describing the
count rate development is given in (a).

For the spectral analysis, best-fit values of some key parameters are shown: (a) the
count rate of the entire energy band, (b) the inner disk radius $R_{\mathrm{in}}$ and
(c) the temperature $T_{\mathrm{in}}$ from the \texttt{diskbb} model. Here, we derived
$R_{\mathrm{in}}$ from $(R_{\mathrm{in}}/D_{10})^2 \cos{(\theta)}$, where $D_{10}=0.3$
is the distance in the 10~kpc unit \citep{gandhi_gaia_2019,Atri2020} and
$\theta=70$~deg is the inclination angle \citep{Torres2019}. No model fitting
is performed in the rise phase due to rapid spectral changes within the duration for the
fitting.

For the temporal analysis, (d) the correlation at the zero lag in the CCF ($r_0$), (e)
the absolute value of the hard lag at the dCCF peak ($|\tau_{\mathrm{hard}}|$), and (f)
the one for soft ($|\tau_{\mathrm{soft}}|$) are plotted. The $|\tau_{\mathrm{soft}}|$
values are quantized at 10 ms due to the statistical limit. Since all the energy band
pairs exhibit similar behavior, we only plot the development of the dCCF between
0.5--1.0~keV and 1.0--2.0~keV.

The overall development of the spectral parameters follows that of the count
rate. Through the plateau and bright decline phases, the count rate change appears
mostly governed by the decreasing $R_{\mathrm{in}}$. In the hard-soft transition phase,
it is difficult to distinguish which of the $T_{\mathrm{in}}$ and $R_{\mathrm{in}}$ are
more responsible for the count rate changes; these two parameters are coupled in the
spectral fitting, which makes the interpretation complicated.
For the temporal parameters of the CCF and dCCF, the development of both the soft and
the hard lags (panels e and f) exhibits a decline in the plateau phase. Beyond the
bright decline phase, though, both lag estimates do not show systematic variations.

\section{Discussion}\label{sec:discussion}
\subsection{Relation between spectral and time-lag components}
\label{ssec:relation_spectral_time-lag}

In \S~\ref{subsec:oneobs}, we dissected the same data set taken at a particular epoch
either in spectral (\S~\ref{subsec:spectrum_fit_1200120106}) and temporal
(\S~\ref{ssubsec:ccf_oneobs}) domain. For the CCF and dCCF, we investigated the energy
dependence (Figures~\ref{fig:CCFDCCF-MULTI_05t1-spec_1200120106_hardlag} and
\ref{fig:CCFDCCF-MULTI_05t1-spec_1200120106_softlag}).

An interesting contrast is observed in the two lags. In the hard lag, the largest
correlation is seen between the 0.5--1.0 and 5--10~keV pairs, where the correlation
increases as the target band is away from the reference band (Figure
\ref{fig:CCFDCCF-MULTI_05t1-spec_1200120106_hardlag} a-2). This suggests that the hard
lag is produced between two spectral components below and above $\sim$3~keV, and that
the harder spectral component becomes {\em more}\/ dominant at higher energies. In the
soft lag, in contrast, the largest correlation is seen between the 0.5--1.0 and
1.0--2.0~keV bands, where the correlation decreases as the target band is away from the
reference band (Figure \ref{fig:CCFDCCF-MULTI_05t1-spec_1200120106_hardlag} a-2). This
suggests that the soft lag is produced between two components, one of which is localized
below 1~keV and the other extending from below 1~keV up to above 5~keV.

In combination with the spectral result, we interpret that the hard lag is between the
disk black body component and the Comptonized component, while the soft lag is between
the Comptonized component and the soft excess component. We verify our interpretation
with simulation. The goal is to examine if the observed features in the dCCF
(Figures~\ref{fig:CCFDCCF-MULTI_05t1-spec_1200120106_hardlag} and
\ref{fig:CCFDCCF-MULTI_05t1-spec_1200120106_softlag}) can be reproduced from
synthesized light curves of each energy band with the mixing ratio of
the three spectral components fixed to the ratio derived from the spectral fitting
(\S~\ref{subsec:spectrum_fit_1200120106}).

\begin{figure}
 \plotone{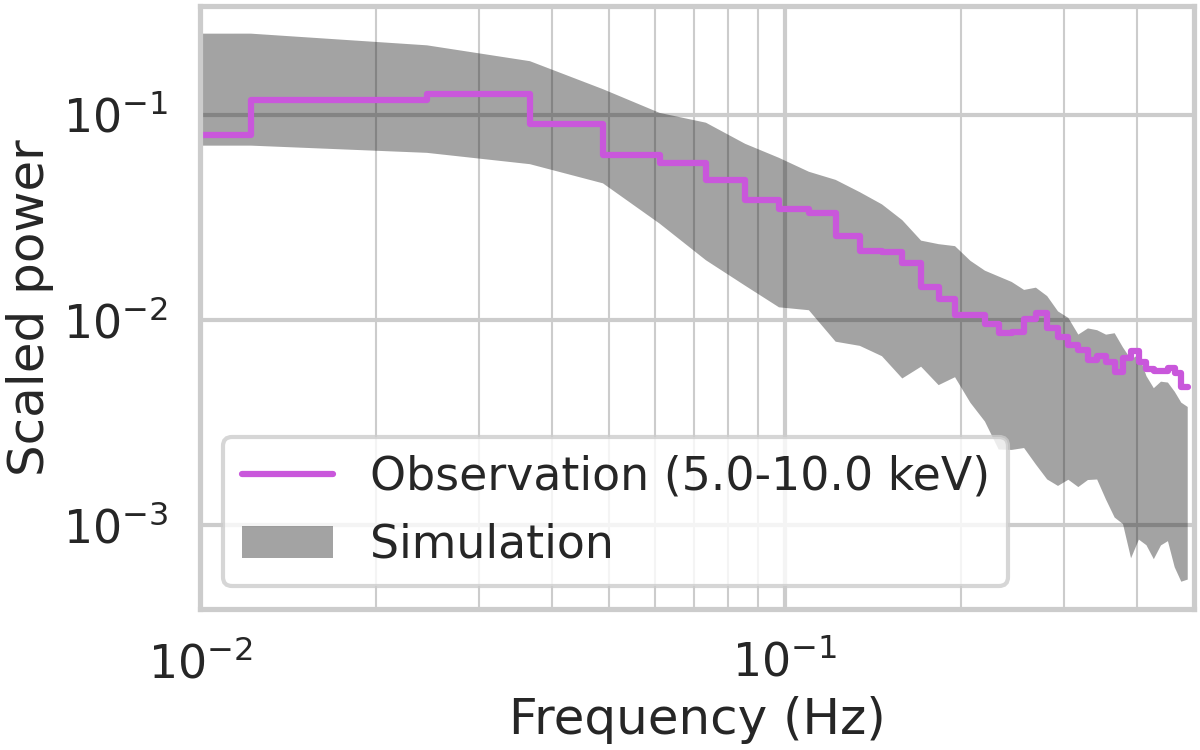}
 \caption{PSDs in the 5.0--10.0~keV, where the Comptonized component is dominant, for
 the observation (purple) and the range of the simulation (gray) with the 80~\%
 confidence intervals derived from 100 realizations. A second component
 is necessary in the simulation to account for the observed PSD at the high frequency
 end, but was omitted as the slight deviation does not affect the result.}
 \label{fig:psd_dt1e-2_en5-10keV}
\end{figure}

To generate the synthesized light curves, we constructed the power spectral density (PSD) of
J1820 using the observation. Figure~\ref{fig:psd_dt1e-2_en5-10keV} shows the PSD for the
observed light curve in the 5.0--10.0~keV, in which the emission is almost purely made
of the Comptonized component (Figure~\ref{fig:XSPEC-FIT_1200120106}). As is often
seen in BHBs, the PSD is represented by a broken power law, in which the power of the
PSD as a function of the frequency breaks at a cut-off frequency. In
Figure~\ref{fig:psd_dt1e-2_en5-10keV}, the cut-off is around $0.02$~Hz.

The Ornstein-Ulenbeck (OU) process is suitable for generating
synthesized light curves having a
PSD of a broken power law shape and was applied to model light curves of BHBs
\citep{kelly_stochastic_2011}. The autocovariance function is expressed as
\begin{equation}
 R_{\rm OU}(t) = A e^{-|t|/t_{\text{scale}}},
\end{equation}
where $A$ is the amplitude of the variance and $t_{\text{scale}}$ is the timescale of
the variation, or the inverse of the cut-off frequency. Computing the Fourier transform
of the auto-covariance, we derive the PSD of the OU process as
\begin{equation}
 \begin{split}
 P_{\rm OU}(\omega)
 &= \frac{1}{2\pi} \int^\infty_{-\infty} e^{-i\omega t} R_{\rm OU}(t) dt \\
 &= B\frac{1}{\omega^2 + \omega_0^2},
 \end{split}
\end{equation}
where $B = A/2\pi \, t_{\text{scale}}$ and $\omega_0 = 1/t_{\text{scale}}$. The PSD
decays as $1/\omega^2$ for $\omega \gg \omega_0$.

\begin{figure}
 \plotone{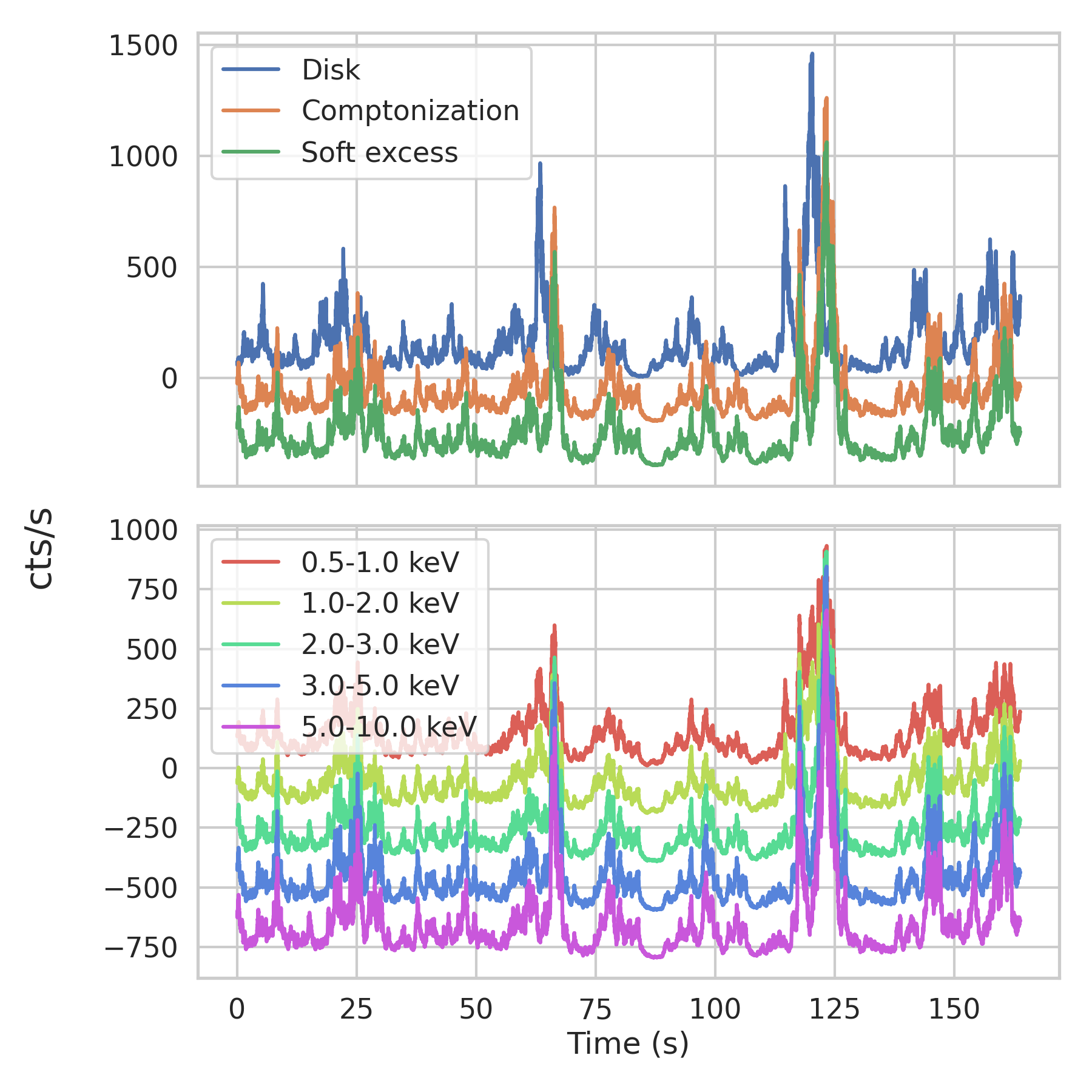}
 \caption{ 
 Top: The synthesized light curves} of the three spectral components, where offsets of --200 and
 --400 counts/s are added to the Compton scattering and soft-excess light curves for
 visibility.
 Bottom: The blended light curves in each energy band for one out of the 100
 samples. Offsets of 0, --200, --400, --600, and --800 counts/s are added for
 0.5-1.0 keV, 1.0-2.0 keV, 2.0-3.0 keV, 3.0-5.0 keV, and 5.0-10.0 keV light curves to
 highlight differences in variability.
 \label{fig:CCFDCCF-SIM_hardlag-softlag_1200120106_original-blended-curves}
\end{figure}

Figure~\ref{fig:CCFDCCF-SIM_hardlag-softlag_1200120106_original-blended-curves} (top) shows
the generated light curves for the three spectral components with 
$t_{\mathrm{scale}}=$5~s (=$1/0.02$~Hz), a time bin of 0.01~s, and a duration of
163.84~s (=$2^{14}$ bins). Here, among the three spectral components, the Comptonized
component is delayed from the disk component by 3~s and the soft excess component is
delayed from the Comptonized component by 0.03~s. The observed delays have some spread as can be
found in the dCCF (Figures~\ref{fig:CCFDCCF-MULTI_05t1-spec_1200120106_hardlag} and
\ref{fig:CCFDCCF-MULTI_05t1-spec_1200120106_softlag}), but we represented them with a
delta function for simplicity. We linearly added them with the ratio derived from the
spectral fitting (\S~\ref{subsec:spectrum_fit_1200120106}) to generate the light curves
in the five fine energy bands. Then, the CCF and dCCF were calculated in the same way as
in \S~\ref{ssubsec:ccf_oneobs} and presented in the same way as in
Figures~\ref{fig:CCFDCCF-MULTI_05t1-spec_1200120106_hardlag} and
\ref{fig:CCFDCCF-MULTI_05t1-spec_1200120106_softlag}.

\begin{figure*}
 \plotone{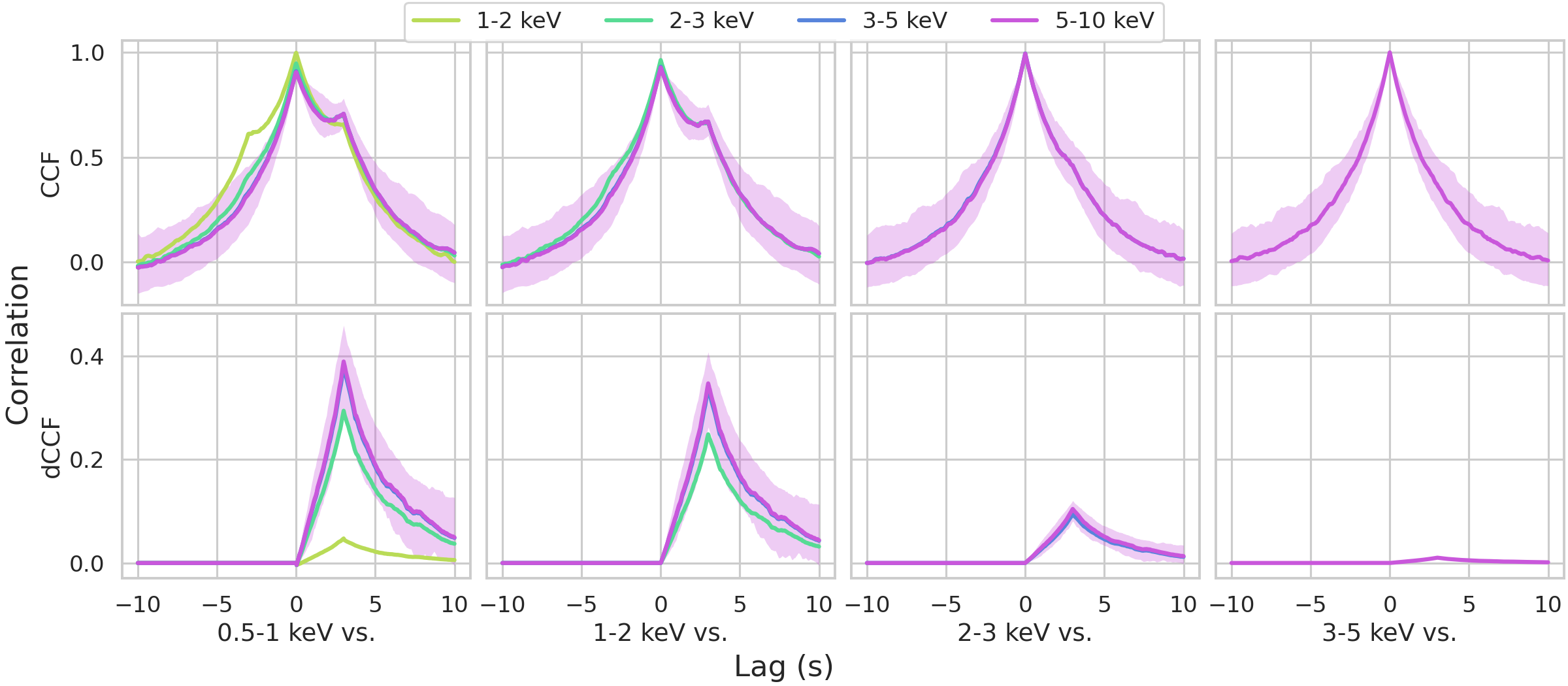}
 \caption{Same as Figure~\ref{fig:CCFDCCF-MULTI_05t1-spec_1200120106_hardlag} but for
 the simulated light curves in
 Figure~\ref{fig:CCFDCCF-SIM_hardlag-softlag_1200120106_original-blended-curves}.
 The 80~\% confidential ranges are shown 
only for the target band 
 5-10~keV.
 }
 \label{fig:CCFDCCF-SIM_hardlag_1200120106}
\end{figure*}

\begin{figure*}
 \plotone{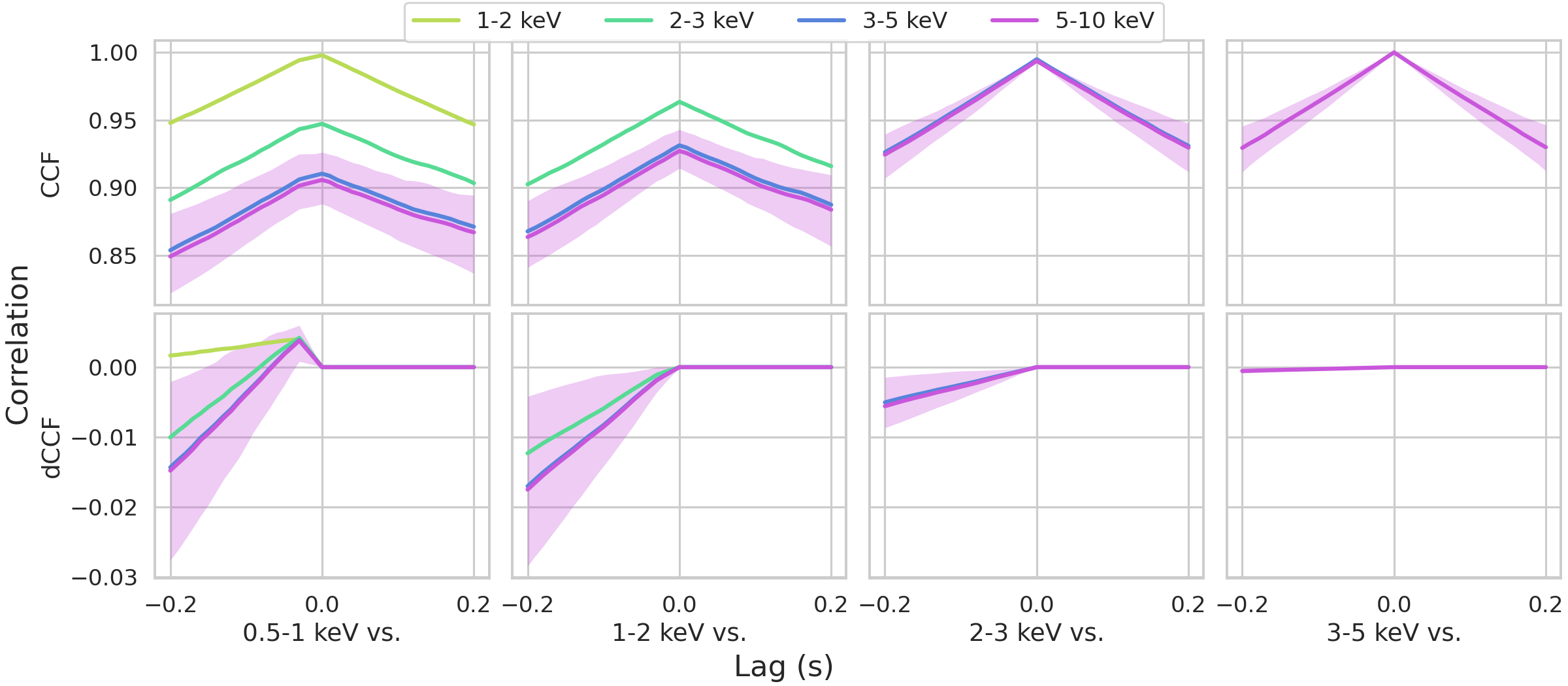}
\caption{Same as Figure~\ref{fig:CCFDCCF-MULTI_05t1-spec_1200120106_softlag} but for
 the simulated light curves in
 Figure~\ref{fig:CCFDCCF-SIM_hardlag-softlag_1200120106_original-blended-curves}.}
\label{fig:CCFDCCF-SIM_softlag_1200120106}
\end{figure*}

Figures~\ref{fig:CCFDCCF-SIM_hardlag_1200120106} and
\ref{fig:CCFDCCF-SIM_softlag_1200120106} show the result of the simulation, where their
energy dependence is very similar to the observed one. In the hard lag, the correlation
increases with the harder target band against the reference band.
In the soft lag, a positive lag was found between the 0.5--1.0~keV and 1.0--2.0~keV
bands.
All these features
are consistent with properties derived from the observed data, which reinforces our
interpretation relating the spectral and temporal features.

\medskip

 We should note that the Comptonized component alone can exhibit an energy-dependent
in its ACF \citep[e.g.][]{poutanen1999spectral,Kotov2001,Arevalo2006a,mahmoud2018physical,veledina2018interplay},
thus yields a residual in the dCCF as calculated in \S~\ref{ssubsec:ccf_oneobs}. However,
if this is what we observe, the CCF asymmetry should be the strongest in the 3--5 versus
5--10~keV CCF, at which the Comptonized components are most
dominant. Figure~\ref{fig:CCFDCCF-MULTI_05t1-spec_1200120106_hardlag} (d--2) shows the
contrary. One could further argue for multiple Comptonization components making the
observed dCCF features due to energy-dependent lags between them, but there is no need
to add a second Comptonization component in the spectral analysis presented in
\S~\ref{subsec:spectrum_fit_1200120106}, thus we do not favor such
interpretation.

\subsection{Interpretation for the parameter evolution}\label{subsec:intepretation_for_parameter_evolution} 
Based on the relation between the spectral and temporal properties
(\S~\ref{ssec:relation_spectral_time-lag}), we now interpret the results
along with their evolution (Figure~\ref{fig:ts_parameters}) for the soft
(\S~\ref{soft_lag}) and hard (\S~\ref{hard_lag}) lags.

\subsubsection{Soft lag}
\label{soft_lag} The soft lag was interpreted as the reverberation lag, in which the
Comptonized photons in the corona irradiate the accretion disk to produce thermal soft
excess emission. Both \citet{Kara2019} and \citet{DeMarco2021} assessed the
soft-lag amplitude by focusing on the different aspects of the cross spectrum.
\citet{Kara2019} stated that the lag amplitude is short based on the phase
shifts in the cross spectrum, which is subject to the underestimate due to the spectral
dilution. \citet{DeMarco2021} alleviated this effect by using the zero-crossing point in
the cross spectra, but it was difficult to eliminate the effects of the overlapping hard
lag signals.

It is thus important to check these results with the complimentary CCF. The dCCF
technique has the advantage of being free from the spectral dilution and separating the
two different time lags. Its estimated soft lag amplitude is $\sim$30~ms on 2018 March
21 (\S~\ref{ssubsec:ccf_oneobs}), which is closer to the value by \citet{DeMarco2021}.
The value corresponds to the light-crossing scale of $\sim$300 times the 
Schwarzschild radius ($R_{\mathrm{s}}$)
for a 10~$M_{\odot}$ black hole, which is more than an order larger than the value
to support the lamp post configuration \citep{Kara2019}. 
If we use the soft lag amplitude derived by the dCCF technique, the
reverberation distance changes from 300 to 100 $R_{\mathrm{s}}$ through the plateau
phase (Figure~\ref{fig:ts_parameters} panel f). Both the spectral and temporal
assessments have systematic uncertainty in the absolute valuesof $R_{\mathrm{in}}$, but
their fractional changes match well(Figure~\ref{fig:ts_parameters} panel b). These are
in line with the picture that the inner radius of the truncated accretion disk kept
contracting through the phase.

\subsubsection{Hard lag}
\label{hard_lag}
\begin{figure}
 \plotone{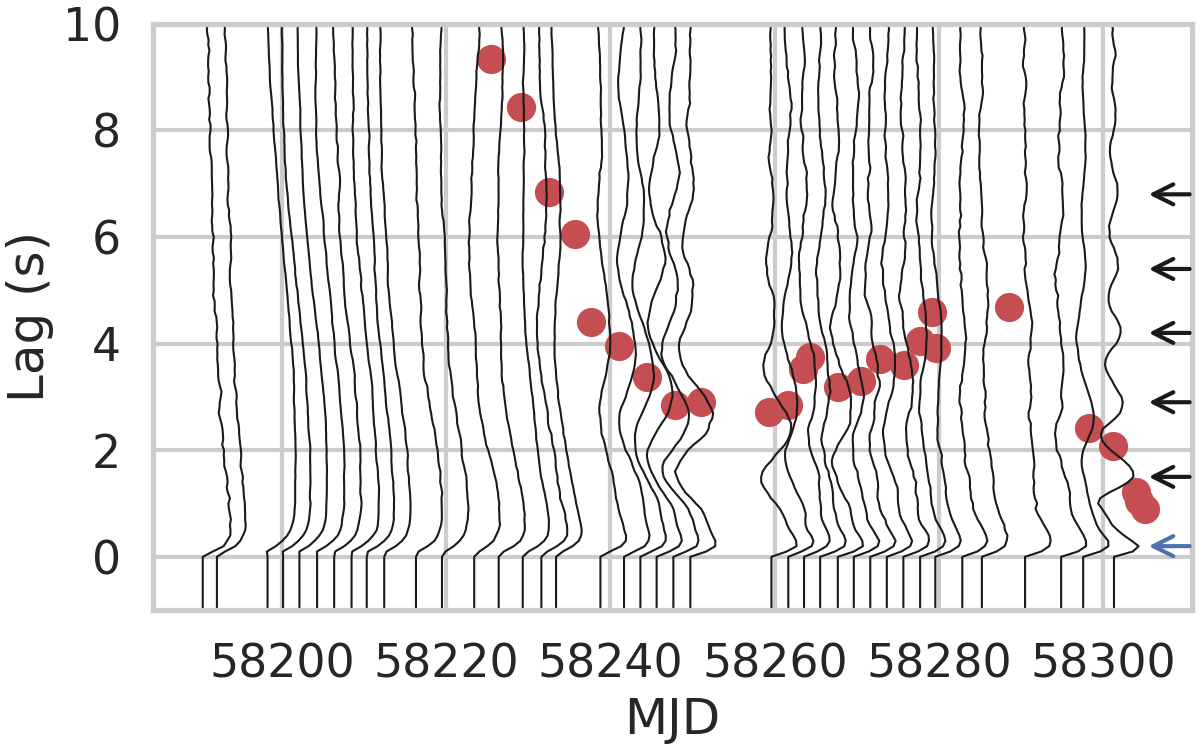}
 \caption{
 Development of the dCCF shown in the sideways in black curves, where often there are several
 peaks in the lags. For example, in the right-most dCCF, the primary (blue arrow) and
 higher-order (black arrow) peaks are marked. The primary is equivalent to
 $|\tau_{\mathrm{hard}}|$ in the panel (e) in Figure~\ref{fig:ts_parameters}. The QPO
 periods in \citet{Stiele2020} added to the primary
 peaks ($|\tau_{\mathrm{hard}}|+1/\nu_{\mathrm{QPO}}$) are shown in
 red circles. These match well with the secondary peaks.
 }
 \label{fig:waterfallplot_dccf_with_qpo}
\end{figure}

The hard lag is a common feature in Galactic BHBs \citep{Uttley2014} and was indeed
found in J1820 too. The lag amplitude is too large for reverberation distances and is
interpreted as a kind of  propagation delay through the accretion disk flow
\citep{Kotov2001,Arevalo2006a}. \citet{kawamura_full_2022} successfully explained the
observed hard lag,
along with the X-ray spectrum and the PSD, by introducing a radial structure in the
accretion corona flow from the inner radius of the accretion disk to the black hole. We
interpret that the hard lag is produced between the disk emission and the Comptonized
emission in the corona (\S~\ref{ssec:relation_spectral_time-lag}). The steady decrease
of the hard lag amplitude ($|\tau_{\mathrm{hard}}|$) in the plateau phase
(Figure~\ref{fig:ts_parameters} panel e) is consistent with the contracting system,

We point out that the dCCF representation of the hard lag shows an association with the
quasi-periodic oscillation (QPO). \citet{Stiele2020} presented the evolution of the QPO
fundamental frequency ($\nu_{\mathrm{QPO}}$) from 0.03 to 10~Hz in the duration of
interest in this paper in their Figure 8. This QPO signal is also seen in the time
domain in the dCCF. Figure~\ref{fig:waterfallplot_dccf_with_qpo} shows time series of
the dCCF at different epochs represented sideways. Close to the 0~s in the lag, the main
peak is recognized in every dCCF, which is the one shown in panel (e) in
Figure~\ref{fig:ts_parameters}. In addition, the secondary (and even higher orders in
some) peaks are found. We have found that the time separation between the main peak and
the secondary peak matches well with the QPO period ($1/\nu_{\mathrm{QPO}}$).
This strongly suggests that the hard lag and the QPO are closely related, and both
originate in the propagation in the accretion flow.

\section{Summary and Conclusion}\label{sec:conclusion}
We explored the NICER X-ray data of MAXI J1820$+$070, a transient BHB, to
analyze time lags between two different X-ray bands during the first 120 days of
discovery when the dense NICER coverage was made. We employed a time-domain
approach (CCF) as opposed to the often used frequency domain
approach (cross spectrum). We
constructed the CCF between the soft (0.5--1.0 keV) and hard (1.0--10 keV) energy bands
and removed the dominating autocorrelation component by subtracting the negative
part of the CCF from the positive part, or vice versa, to derive the differential CCF
(dCCF). In the dCCF, we clearly identified both the soft and hard lags respectively of
$\sim$0.03 and $\sim$3~s separately without being diluted by the spectral mixture. The
values are closer to the ones derived from the cross spectra analysis in
\citet{DeMarco2021} than those in \cite{Kara2019}. This demonstrates the effectiveness
of the dCCF complementary to the cross-spectrum in the X-ray time lag
analysis, particularly when combined with the rich statistics that NICER
brings.

We conducted the spectral and timing (dCCF) analyses separately at one of the brightest
epochs in the duration of interest. In the spectral analysis, the X-ray spectra are
represented by three components; the Comptonized power-law emission, the disk black body
emission, and the soft excess emission (\S~\ref{subsec:spectrum_fit_1200120106}).  In
the timing analysis, we derived both the soft and the hard lags in the dCCF and
constrained the energy dependence of the correlation by using a finer energy band
resolution (\S~\ref{ssubsec:ccf_oneobs}). We then applied the analysis to the entire
duration of interest to track the development of both the spectral and timing properties
(\S~\ref{sec:ccf_time-development}).

Based on the spectral and timing analysis results, we conjectured that the soft lag is
between the Comptonized and the soft excess emission, while the hard lag is between the
disk and Comptonized emission. We verified this hypothesis by constructing the dCCF
based on synthesized light curves of different energy bands made as a
linear composition of the three spectral components lagged from each
other. The energy dependence of the dCCF was recovered, which supports our hypothesis
(\S~\ref{ssec:relation_spectral_time-lag}).

Finally, we investigated the evolution of the spectral and timing properties along with
time. Both the soft and the hard lags kept decreasing through the plateau phase. In the
following bright decline phase, we could not follow the lag evolution due to the limited
statistics.  The evolution of the hard and soft lag amplitude is in agreement with the
one of $R_{\mathrm{in}}$.  All these features are consistent with a picture that the
accretion disk is truncated and its inner radius kept contracting through the plateau
phase (\S~\ref{soft_lag}). Finally, we report a clear link between the hard lag and the
QPO in the time domain (\S~\ref{hard_lag}).

\begin{acknowledgments}
This work was supported by JSPS KAKENHI Grant Number 22J13440.
The authors appreciate useful discussions with Tenyo Kawamura at IPMU.
M.\,M.\ is financially supported by the Hakubi project of Kyoto University.
\end{acknowledgments}

\facilities{NICER}
\software{
 XSPEC 12.12.1         \citep{arnaud1996xspec},
 HEAsoft 6.30         \citep{blackburn1995ftools,nasa_high_energy_astrophysics_science_archive_research_center_heasarc_heasoft_2014},
 astropy 5.1          \citep{robitaille_astropy_2013,collaboration_astropy_2018,collaboration_astropy_2022},
 matplotlib 3.5.2       \citep{hunter_matplotlib_2007}
 numpy 1.22.3         \citep{harris_array_2020},
 scipy 1.7.1          \citep{virtanen_scipy_2020}
 tensorflow 0.14.0 \citep{abadi_tensorflow_2016}
}


\bibliography{refs}{}
\bibliographystyle{aasjournal}
\end{document}